\pdfoutput=1

\documentclass[11pt,final]{article}

\usepackage[preprint]{acl}
\usepackage{xspace}
\usepackage{booktabs}
\usepackage{adjustbox}
\usepackage{acronym}
\usepackage{flushend}
\usepackage{graphicx}
\usepackage[htt]{hyphenat}
\usepackage[]{xcolor}
\usepackage{tabularx}
\usepackage{multirow}
\usepackage{subcaption}
\usepackage{arydshln}
\usepackage{amsmath} 
\usepackage[inline]{enumitem}
\usepackage{anyfontsize}
\definecolor{color1}{rgb}{0.2, 0.81, 0.30}
\definecolor{color2}{rgb}{0.66, 0.25,0.97}
\definecolor{color3}{rgb}{0.92, 0.4, 0.08}
\definecolor{color4}{rgb}{0.44, 0.4, 0.9}

\newcommand{\cast}{{\ac{CAsT}}\xspace}
\newcommand{\ikat}{{\ac{iKAT}}\xspace}

\newcommand{\OurModel}{MQ4CS\xspace}
\newcommand{\AQD}{{{MQ4CS$_\text{ans}$}}\xspace}

\newcommand{\QD}{{{MQ4CS}}\xspace}
\newcommand{\AD}{{{AQ}}\xspace}
\newcommand{\topiocqa}{{{TopiOCQA}}\xspace}
\newcommand{\AQDAnswer}{{{MQ4CS$_\text{ans}$+rerank}}\xspace}

\newcommand{\QR}{{{QR}}\xspace}
\newcommand{\llama}{{{LlaMA}}\xspace}
\newcommand{\gptF}{{{GPT-4}}\xspace}

\newcommand{\llamaQR}{{{LlaMAQR}}\xspace}
\newcommand{\ConvGQR}{{{ConvGQR}}\xspace}

 \newcommand{\ourdata}{MASQ\xspace}

\newcommand{\header}[1]{\vspace*{2mm}\noindent\textbf{#1}}

\acrodef{CIS}{conversational information seeking}
\acrodef{CS}{conversational search}
\acrodef{IR}{information retrieval}
\acrodef{NLP}{natural language processing}
\acrodef{PLM}{pretrained language model}
\acrodef{LLM}{large language model}
\acrodef{TREC}{TExt Retrieval Conference}
\acrodef{CAsT}{Conversational Assistance Track}
\acrodef{iKAT}{Interactive Knowledge Assistance Track}
\acrodef{RG}{retrieve-then-generate}
\acrodef{GR}{generate-then-retrive}
\acrodef{RAG}{retrieval-augmented generation}
\acrodef{GAR}{generation-augmented retrieval}

\allowdisplaybreaks

\DeclareMathOperator*{\argmax}{arg\,max}

\begin{document}

\title{Leveraging LLMs as Multi-Aspect Query Generators for Conversational Search}
\title{Generating Multi-Aspect Queries for Conversational Search}

\author{Zahra Abbasiantaeb \\
  University of Amsterdam \\
  Amsterdam \\
  The Netherlands \\
  \texttt{z.abbasiantaeb@uva.nl} \\\And
  Simon Lupart \\
  University of Amsterdam \\
  Amsterdam \\
  The Netherlands \\
  \texttt{s.c.lupart@uva.nl} \\\And
  Mohammad Aliannejadi \\
  University of Amsterdam \\
  Amsterdam \\
  The Netherlands \\
  \texttt{m.aliannejadi@uva.nl} \\}


\maketitle

\begin{abstract}
\Acf{CIS} systems aim to model the user's information need within the conversational context and retrieve the relevant information. One major approach to modeling the conversational context aims to rewrite the user utterance in the conversation to represent the information need independently.
Recent work has shown the benefit of expanding the rewritten utterance with relevant terms. In this work, we hypothesize that breaking down the information of an utterance into multi-aspect rewritten queries can lead to more effective retrieval performance. This is more evident in more complex utterances that require gathering evidence from various information sources, where a single query rewrite or query representation cannot capture the complexity of the utterance. 
To test this hypothesis, we conduct extensive experiments on five widely used \ac{CIS} datasets where we leverage LLMs to generate multi-aspect queries to represent the information need for each utterance in multiple query rewrites. We show that, for most of the utterances, the same retrieval model would perform better with more than one rewritten query by 85\% in terms of nDCG@3. 
We further propose a multi-aspect query generation and retrieval framework, called \QD.
Our extensive experiments show that \QD outperforms the state-of-the-art query rewriting methods.
We make our code and our new dataset of generated multi-aspect queries publicly available.

\end{abstract}

\section{Introduction}

\begin{figure}
    \centering
    \includegraphics[width=0.5\textwidth]{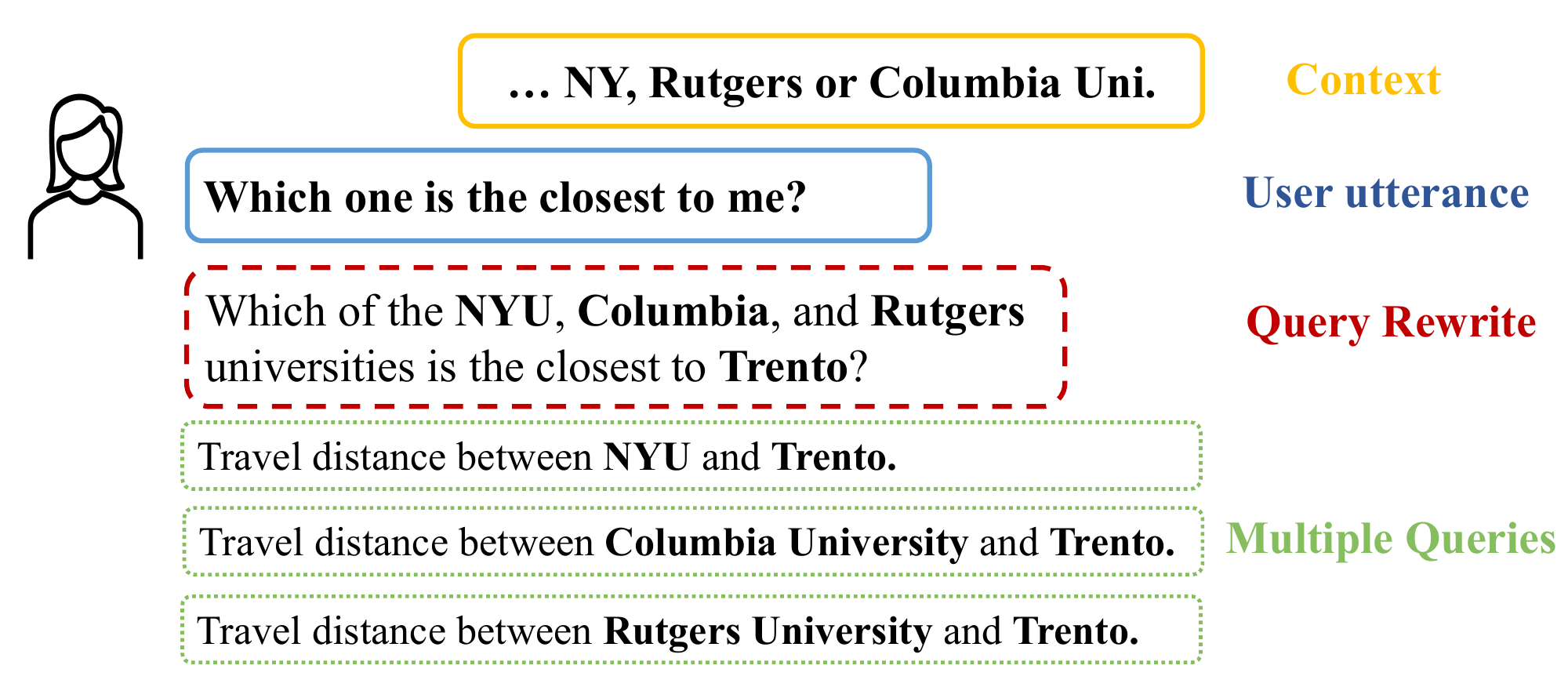}
    \caption{An example conversation with a complex user utterance. The system needs to generate three distinct queries and search for every query.}
    \label{fig:example}
\end{figure}

\Acf{CIS} is a well-established topic in \acf{IR}~\cite{zamani2022conversational,aliannejadi2024interactions}, where a knowledge assistant interacts with the user to fulfill their information needs. While conversations can be complex~\cite{radlinski2017theoretical}, involving various types of interactions such as revealment and clarification, one of the main goals of the system is to provide an answer to the user's request. As a conversation is prolonged, several challenges and complexities arise, such as language dependency (e.g., anaphora, ellipsis), long conversation context modeling, and more complex information needs~\cite{dalton2020trec,owoicho2022trec,aliannejadi2024trec}. Much research aims to address these issues by utterance rewriting, where the written query is aimed to resolve language dependencies and encapsulate the context and information complexities~\cite{Voskarides2020query,yu2021few}. 

Encapsulating the complex nature of the conversational information need in one single rewritten utterance can lead to several limitations, especially in cases where the query cannot be answered using a single passage and requires complex reasoning over multiple facts from different sources in a chain-of-thought scenario~\cite{aliannejadi2024trec,lyu2023faithful}.
Take the user query of Figure~\ref{fig:example} as an example. It is unlikely to find a passage that has distance information about all these universities compared to the user's address. Therefore, the system would need to gather relevant information from different sources (e.g., distance of each city) and reason over the gathered evidence to generate the final response. Existing ranking methods rely solely on semantic similarity between a query and a passage, without high-level reasoning or control over the set of retrieved passages. For example, they do not ensure that the top results contain address information about all three universities the user is interested in. Therefore, there is no guarantee that the passages containing different pieces of relevant information would be ranked high. 

Recent research suggests that retrieving passages based on multiple queries can lead to improved retrieval performance~\cite{mao-etal-2023-large,kostric2024surprisingly}.
Existing methods like LLM4CS, however, focus on prompting the LLM multiple times using the same prompt to mitigate the LLM's bias in generating a single query rewrite.
To the best of our knowledge, no research has systematically studied and analyzed the impact of breaking information need into multi-aspect queries. 
To address this research gap, we hypothesize that the majority of conversational information needs cannot be summarized into a single query rewrite, as the conversation evolves and the user's information need becomes more complex. To test our hypothesis, in a single LLM call, 
we prompt the \ac{LLM} to generate 1--5 multi-aspect queries for each utterance for five major \ac{CIS} datasets and measure the performance of the same retrieval and reranking pipeline. Assuming that the system knows the optimum number of generated queries based on an Oracle setting, we observe that more than 65\% of the utterances exhibit better performance using the multi-aspect generated queries,
leading to 85\% performance improvement in nDCG@3. This verifies our hypothesis, showing that the majority of the utterances benefit from multi-aspect query generation. Based on these findings, we build a new dataset, called \ourdata, consisting of multi-aspect query rewrites for each user utterance, together with the optimum number of queries to represent a user utterance.

Inspired by our findings, we propose a simple yet effective conversational retrieval framework based on generating multi-aspect queries, called \OurModel (\textbf{Mu}lti-aspect \textbf{Q}uery Generation and Retrieval \textbf{for} \textbf{C}onversational \textbf{S}earch). \OurModel takes a conversational utterance as input and generates a given number of queries to address the information need from multiple aspects. It then retrieves and reranks passages for each query and in the final step does rank list fusion to output a single ranking. \OurModel relies on the \ac{LLM} internal knowledge and reasoning capacity to model the user's information need and generate multi-aspect queries. 

We conduct extensive experiments on five widely used \ac{CIS} datasets, namely, \ac{TREC} \cast 19, 20, \& 22, \ac{TREC} \ikat 23, and \topiocqa. We show that \OurModel outperforms the SOTA query rewriting approaches on all the datasets in terms of various metrics. 
Furthermore, we study the effect of multi-aspect query generation in various experimental setups, such as different levels of complexity, and topic switches. We observe a higher performance gap between \OurModel and SOTA models as the complexity of user utterance increases, demonstrating the effectiveness of multi-aspect query generation for complex user queries.

We summarize our contributions as follows:
\begin{itemize}[nosep,leftmargin=*]
    \item We propose a conversational passage retrieval framework by leveraging the \ac{LLM}'s internal knowledge to rewrite user utterances to multi-aspect queries and fuse their rankings.
    \item We show generating multi-aspect queries improves retrieval performance for the majority of queries. To facilitate research in this area and establish it as a new task, we build and release a multi-aspect query dataset, called \ourdata, focusing on five major conversational search datasets.\footnote{The dataset is provided in the supplementary materials and will be made publicly available upon acceptance.}
    \item We conduct extensive experiments, showcasing the effectiveness of \OurModel for conversational passage retrieval on five major conversational search datasets using commercial and open-source \acp{LLM}.
\end{itemize}

\section{Related Work}

Recently, \ac{CIS} has gained significant popularity in both \ac{IR} and \ac{NLP} communities~\cite{anand2020conversational}. Similar to knowledge-intensive dialogues~\cite{dinan2019wizard,feng2021multidoc,li2022knowledge}, a key challenge in \ac{CIS} is to model the dialogue context to better understand the user information need and perform effective retrieval~\cite{zamani2022conversational}.
\ac{TREC} \cast 19--22~\cite{dalton2020trec} and \ikat 23~\cite{aliannejadi2024trec} aim to address these challenges through a common evaluation framework in which complex and knowledge-intensive dialogues were provided to participants, as well as several passage collections. The goal was to retrieve relevant passages for each turn in a dialogue and generate a response synthesizing several passages. 
\ac{TREC} \cast 22~\cite{owoicho2022trec} advanced this track by introducing mixed-initiative (e.g., clarifying questions~\cite{rao2018learning,aliannejadi2019asking}) and user feedback~\cite{owoicho2023exploiting} turns, as well as going beyond single conversation trajectory for a given topic. \ac{TREC} \ikat 23 focuses on the long-term personal conversational memory of the model via introducing a personal knowledge graph.

A line of research aims at learning to represent the dialogue context directly for passage retrieval~\cite{yu2021few,hai2023cosplade}, where a distillation loss learns to map the representation of the whole dialogue context to the one of the gold resolved query, hence improving the dense retrieval performance. The INSTRUCTOR~\cite{jin-etal-2023-instructor} model trains the document encoder model by using the relevance score predicted by LLMs. 

Most existing methods tackle the context modeling problem by query rewriting where the goal is to address the ambiguity and dependence of a user utterance by resolving its dependencies and making it self-contained~\cite{voskarides2019,Voskarides2020query,lin2021}. The rewritten query is supposed to be a self-contained and context-independent query that represents the user's information needs per turn. 
CRDR~\cite{qian2022} forms the rewritten query by modifying the query by disambiguation of the anaphora and ellipsis.
The existing work trains GPT-2~\cite{yu2020few,vakulenko2021question} and T5~\cite{dalton2020trec} models on the CANARD dataset~\cite{elgohary2019} to generate the rewritten query. 
LeCoRE~\cite{Mao:2023-Lecore} is an extension of the SPLADE model~\cite{formal2021splade} obtained by denoising the representation of the context. The denoising model works by distilling knowledge from query rewrite.
ConvGQR~\cite{mo-etal-2023-convgqr} model expands the query rewrite with potential answers. They train two separate models for the query rewrite and answer generation. CONQRR~\citep{wu-etal-2022-conqrr} trains the T5 model using reinforcement learning to generate query rewrite based on the retrieval performance and achieves a better performance compared to the T5QR~\cite{raffel2020exploring} 
model.
\citet{ye-etal-2023-enhancing} propose using LLMs as zero- and few-shot learners in two steps including query rewriting and rewrite editing to form the query rewrite. 
LLM4CS~\cite{mao-etal-2023-large} employs different prompting strategies and creates multiple query rewrites and answers. The embedding of query rewrites and answers are combined using various methods and the aggregated representation is used for retrieval. 
LLM4CS is the most similar work to ours. Our work distinguishes itself from LLM4CS in various aspects:
\begin{enumerate*}[label=(\roman*)]
    \item LLM4CS does not prompt the LLM to generate multi-aspect queries. Instead, it prompts for a query rewrite and repeats this prompt five times to get different generation variations, with no guarantee that the generated queries address different aspects of the original query. \QD, instead prompts the LLM to break the user information need into multiple queries by generating various multi-aspect queries, focusing on different perspectives.
    \item LLM4CS either selects one of the five generated queries or computes the average of representations of multiple queries for retrieval. Therefore, the retrieval task is only done based on one query/representation. \QD, on the other hand, aims not to miss any documents that can be retrieved by each single query. Therefore, it does passage retrieval for all five queries independently and fuses their rankings.
\end{enumerate*}

\section{Methodology}
\subsection{Task Definition}
Each conversation includes several turns, where a turn starts with a user utterance $u_i$, followed by a system response $r_i$. The context of the conversation at turn $i$ is represented as $c_i=\{(u_1, r_1),..., (u_{i-1}, r_{i-1})\}$. Different from other datasets, \ac{TREC} \ikat 23 also includes the persona of the user. The persona is a knowledge base, consisting of a set of statements shown as $PTKB = \{s_1,...,s_l\}$.
The task of passage retrieval for conversational assistants is to retrieve relevant passages to the current user utterance $u_i$ from the collection $D = \{d_1, ..., d_{|D|}\}$. The ordered list of retrieved passages for user utterance $u_i$ is shown as $D^{\prime}_{i}$ which is a subset of $D$.

\begin{table*}[t]
\centering
\caption{Passage retrieval performance of the proposed models and baselines on TREC \cast 20 \& 22 and \ikat 23 datasets. The best results that are significantly (t-test $p\_value \leq 0.05$) better are shown in \textbf{bold}. The second-best results are shown with \underline{underline}. Here we use  $\phi=5$ and \gptF as our \ac{LLM}.}
\adjustbox{max width=1\textwidth}{
 \begin{tabular}{llllllllllll}
\toprule
 \multirow{2}{*}{\textbf{Method}} & \multicolumn{3}{c}{\textbf{\cast 20}} & & \multicolumn{3}{c}{\textbf{\cast 22}} & & \multicolumn{3}{c}{\textbf{\ikat 23}}\\  \cmidrule{2-4} \cmidrule{6-8} \cmidrule{10-12}

& \textbf{nDCG@3} & \textbf{R@100} & \textbf{MRR} & & \textbf{nDCG@3}  & \textbf{R@100} & \textbf{MRR} & & \textbf{nDCG@3}  & \textbf{R@100} & \textbf{MRR}  \\
\cmidrule(lr){1-12} 
GPT4QR   & \textbf{46.8}    & 55.2  & 61.5   && \underline{34.8}   & 30.1  & 52.2    && 21.9    & 22.1  & 34.6  \\
T5QR     & 38.7   & 45.6  & 53.1   && 30.2   & 23.9  & 45.5    && 14.1   & 13.8  & 23.9 \\
ConvGQR  & 35.7   & 47.7  & 49.4   && 25.0    & 22.1  & 41.0   && 14.7    & 13.9  & 21.9\\
\llamaQR  & 36.3  & 49.3  & 51.4    &&  30.8  & 27.3  & 49.0   &&  9.6   & 14.2  & 16.0\\
GPT4-\AD   & \underline{45.2}   & 54.6  & 62.3   && 31.3  & 32.5  & 49.9   && 15.0    & 22.2  & 24.6\\
\llama-\AD   & 19.3  & 31.0  & 29.8    && 21.1  & 21.5  & 34.4    && 10.6  & 18.8  & 16.9 \\
LLM4CS   & 38.7     & 51.1  & 54.5   &&  27.5       &  30.0    &  45.3   &&  10.5       &  14.5    &  16.5  \\
\cmidrule{1-12}
HumanQR    & 50.5   & 61.8  & 66.8   && 41.3   & 37.6  & 60.4   && 30.7   & 35.8  & 43.3 \\
\cmidrule{1-12}
\AQD  & 44.8   & \underline{63.8}  & \textbf{66.5}   && 33.2    & 33.8  & \underline{57.9}   && \textbf{23.0}    & \underline{26.7}  & \underline{38.6} \\
\AQDAnswer & 45.0  & 57.6  & 62.9   && 32.5    & \underline{35.0}  & 51.6   && 18.1     & 25.2  & 30.6\\
\QD   & 44.8  & \textbf{64.8}  & \underline{64.3}   && \textbf{35.0}    & \textbf{36.9}  & \textbf{58.5}   && \underline{22.6}    & \textbf{29.8}  & \textbf{41.9} \\
\cmidrule{1-12}
\multicolumn{12}{c}{\textbf{Oracle}}\\
\cmidrule{1-12}

\AQD   &  63.6   &  71.9  &  79.3   & & 55.6   &  43.7  &  79.9  &&  40.6   &  37.6  &  57.5 \\
\AQDAnswer  &  60.0   &  72.9  &  73.8   & & 56.0     &  43.5  &  81.8 &&  34.2   &  37.4  &  50.8\\
\QD   &  66.0  &  67.8  &  83.7   & & 53.7   &  44.9  &  75.7 && 37.7   &  36.0  &  56.4\\

\bottomrule
\end{tabular}
}
\label{tab:retrieval-results-main-pool-all}

\end{table*}

\subsection{MQ4CS Framework}

\begin{figure}[!htbp]
    \centering
    \includegraphics[width=0.47\textwidth,trim={0.5cm 1.5cm 0.5cm 0.5cm},clip]{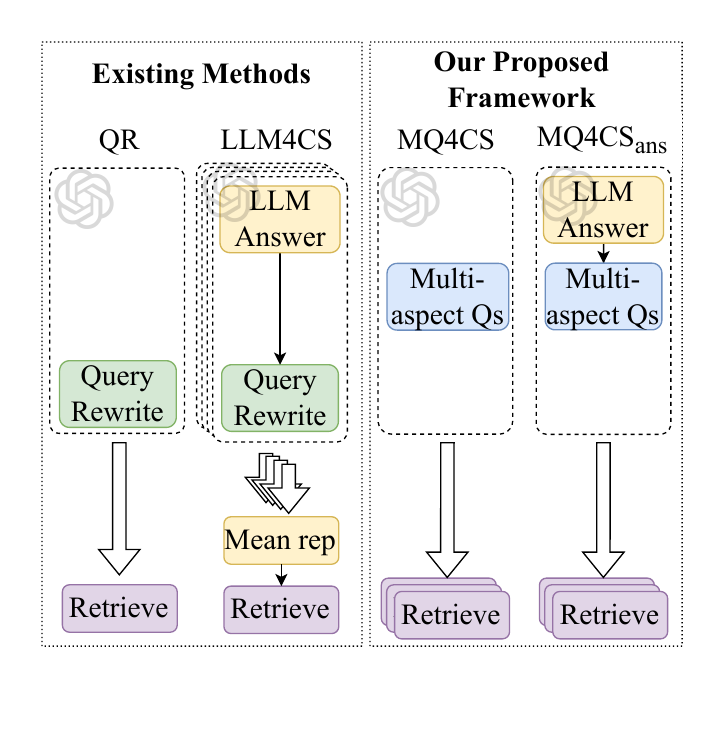}
    \caption{A high-level overview of the proposed framework, compared with existing models. In \QR a single query is generated by LLM and in LLM4CS multiple LLM calls are made to generate different query rewrites. In our \QD and \AQD models, we generate multi-aspect queries in a single prompt. We then perform retrieval on each query independently to avoid information loss.}
    \label{fig:model_arch}
\end{figure}

In this section, we introduce our proposed conversational search framework, called \OurModel (\textbf{M}ulti-aspect \textbf{Q}uery Generation and Retrieval \textbf{for} \textbf{C}onversational \textbf{S}earch). In Figure~\ref{fig:model_arch}, we show the overview of the existing query rewriting techniques (on the left), compared to our proposed framework (on the right). In general, a conversational search framework consists of a query rewriting module that aims to resolve the current utterance's dependencies and generate a stand-alone query. The newly generated query is used for retrieval and reranking. 

Methods like LLM4CS~\cite{mao-etal-2023-large} prompt the LLM multiple times to generate more than one query, and then take the average representation of them to do the ranking.
We take a different approach in the query rewriting phase, that is, we prompt the LLM only once to resolve and generate multi-aspect queries (a maximum of 5) that represent the information need for the current utterance from different aspects (see the example in Figure~\ref{fig:example}). We leverage the internal knowledge and reasoning capabilities of LLMs to understand complex information and break it into multiple queries.  We then pass each query to retrieval and reranking and fuse the final ranking list of each query to output one single ranking for the utterance. We describe our framework's components below. \looseness=-1

\header{LLM answer.}
Inspired by existing work that shows asking LLMs for explanation further improves their performance~\cite{wei2022chain}, as well as the work that shows that conversational search can be improved by asking the LLM to generate an answer~\cite{mao-etal-2023-large}, we ask the LLM to first give a response to the user's utterance.
Therefore, the LLM response generator module instructs the \ac{LLM} to generate the response $r_i^\prime$ to the user utterance given the conversation context. 
Therefore, the generated response in this step could be used by the query generator as shown in Equation~\ref{eq:3}. Our LLM response generation module takes $u_i$, $c_i$, and PTKB (if exists) as input and generates $r^{\prime}_i$, as shown below: \looseness=-1
\begin{equation}
    \label{eq:1}
        r^{\prime}_i = AG (u_i, c_i, \text{PTKB})~.    
\end{equation}

\header{Multi-aspect query generation.}
This module takes the whole conversation context, the user persona (if exists), and the current utterance $u_i$ as input, and prompts an LLM to generate a maximum of $\phi$ queries denoted as $Q^{i} = \{q_1^{i}, ..., q_\phi^{i}\}$ to retrieve passages for user utterance $u_i$. We designed two different prompts for this module for generating multiple queries in one single prompt, with or without the LLM response $r^{\prime}_i$ as input. Therefore, our query generator module looks as below:

\begin{equation}
     \label{eq:3}
    Q^i_{\ \ } = QG (r^{\prime}_i,u_i, c_i, \text{PTKB}, \phi)~,
\end{equation}
\noindent where $\phi$ is the number of queries to generate. We parse the output $Q^i_{\ \ }$ to get the list of generated queries denoted as $\{q_1^{i}, ..., q_{\phi}^{i}\}$.

\header{Retrieval and reranking.} 
Most existing methods follow a two-step approach for retrieval, i.e., first-stage retrieval and reranking~\cite{DBLP:series/synthesis/2021LinNY}. 
We use BM25 from Pyserini \cite{Lin_etal_SIGIR2021_Pyserini} for first-stage retrieval ($Ret$) and the pre-trained Cross-Encoder model \texttt{ms-marco-MiniLM-L-6-v2} from the \texttt{sentence\_transformers} for reranking ($ReRank$). 
This module takes the query $q_k^{i}$ as input and generates $D^{\prime}_{k,i}$ ranked list as output for each query:
\begin{equation}
    \label{eq:4}
        D^{\prime}_{k,i} = ReRank(Ret(D, q_k^{i}),q_k^{i})~.
\end{equation}

\header{Ranked list fusion.}
This module takes the document ranking of each generated query as input (i.e., $\phi$ ranked lists), and produces one final ranking. We follow two simple approaches: (i) interleaving where we simply interleave the $\phi$ ranked lists ($D^{\prime}_{1,i}, ..., D^{\prime}_{\phi,i}$) into one list; (ii) reranking where we use the answer generated by the LLM as an extended query and rerank the passages based on their relevance score to the LLM's response. For interleaving we select the first passage from each list, placing them in the final list according to the order of the queries. Then, we repeat the process with the second passage and so on, while removing the duplicates.

\header{Model variants.}
We propose three variants of our framework. They all follow the same paradigm, with slight variations.

    \begin{itemize}[nosep,leftmargin=*]
        \item \textbf{\QD}: This is our main model that only leverages the query generation (without LLM's answer) and performs ranked list interleaving as fusion. 
            \begin{equation}
            \begin{aligned}
               D^{\prime}_{i} = InterLeave(D^{\prime}_{1,i}, ..., D^{\prime}_{\phi,i})~. \\
            \end{aligned}
        \end{equation}
        \item \textbf{\QD} from \textbf{A}nswer (\textbf{\AQD}): This model leverages the LLM response to generate queries. Therefore it first generates $r^{\prime}_i$ and then pass it to $QG$ module in Equation~\ref{eq:3}. This model also performs ranked list interleaving as fusion.

        \item \textbf{\AQD} reranked with \textbf{A}nswer (\textbf{\AQDAnswer}): 
        This variant uses the LLM response as an extent query to rerank the final ranking. Therefore, the final ranking is produced as follows:        

        \begin{equation*}
            \begin{aligned}
               D^{\prime}_{i} = ReRank(InterLeave(D^{\prime}_{1,i}, ..., D^{\prime}_{\phi,i}),r^{\prime}_i)~. \\
            \end{aligned}
        \end{equation*}
    \end{itemize}

\header{Base LLM.}
As base LLM, we use GPT-4 and GPT-4o. We also report the results using LlaMA 3.1 in Appendix~\ref{sec:different-llm} on page \pageref{sec:different-llm}. 
    
\header{Oracle.} We define an Oracle model that always selects the optimal value of $\phi$ for each user utterance $u_i$ based on the model's performance on the test set, as defined below:  
\[
\phi_i^* = \argmax_{\phi \in \{1,\dots,5\}} QG(r'_i,u_i,c_i,\text{PTKB},\phi)~,
\]
where $\phi_i^*$ denotes the best $\phi$ value of turn $i$ in a conversation. Each user utterance needs a different number of queries depending on the complexity of the information need. The performance of the Oracle model is considered as the upper bound because in this setting, for each user utterance, we issue a different prompt with the given $\phi$.

\header{Prompts.} We use \gptF as a zero-shot learner for the $QG$ module with or without LLM answer. 
The prompts used for these approaches are shown in Tables~\ref{tbl:prompt-aqd-ad-gpt4} and \ref{tbl:prompt-qd-gpt4} on page~\pageref{tbl:prompt-aqd-ad-gpt4}, respectively.
As our preliminary experiments on using the zero-shot prompts for \llama model showed low performance, we progressed into using few-shot prompts for the \llama model. The examples of few-shots are from the pruned turns of the \ikat dataset and predictions of the \gptF model. For answer generation in the $AG$ module, we design a zero-shot prompt. The few-shot prompts designed for $QG$ module with and without LLM answer in \llama model are shown in Tables~\ref{tbl:prompt-aqd-ad-gpt4} and \ref{tbl:prompt-qd-llama} on page~\pageref{tbl:prompt-aqd-ad-gpt4}, respectively.        

\section{Experimental Setup}

We explain our baselines in the following. The datasets, metrics, and hyper-parameters are explained in Appendix \ref{sec:appendix-exp-setup} on page \pageref{sec:appendix-exp-setup}.

\begin{table}[t]
\caption{Passage retrieval results on the \topiocqa dataset with $\phi=5$. The best results that are significantly better (t-tests with $p\_value \leq 0.05$) are in \textbf{bold} face. We use  GPT-4o as our LLM in our proposed framework.}

\centering
\adjustbox{max width=0.5\textwidth}{
 \begin{tabular}{lllll}
\toprule
\textbf{Method} & \textbf{MAP} & \textbf{R@10} &  \textbf{R@100} & \textbf{MRR} \\
\toprule

GPT4oQR  &  45.4  & 66.9  & 80.9  & 45.4      \\
T5QR    &   33.5  & 50.2  & 62.5  & 33.5    \\
\ConvGQR   &  31.1  & 49.0  & 63.2  & 31.1       \\
GPT4o-\AD  &  42.9  & 63.1  & 79.8  & 42.9  \\
LLM4CS & 36.8 & 57.1 & 75.9 & 36.8 \\
\midrule
\AQD  &  \underline{46.7}  & \underline{70.6}  & \underline{87.0}  & \underline{46.7}  \\  
\AQDAnswer & 43.6  & 65.5  & 83.8  & 43.6   \\
\QD   & \textbf{47.5}  & \textbf{72.6}  & \textbf{87.8}  & \textbf{47.5}  \\
\midrule
 \multicolumn{5}{c}{\textbf{Oracle}} \\
\midrule
\AQD  & 58.6  &  79.3  &  91.0  &  58.6   \\
\AQDAnswer &  45.5  &  68.3  &  87.4  &  45.5 \\
\QD       & 59.4  &  80.5  &  91.9  &  59.4   \\

\bottomrule

\end{tabular}
}
\label{tab:retrieval-results-main-pool-topiocqa}

\end{table}

\header{Compared methods.} We compare our proposed models to five strong query rewriting (QR) baselines including (1) \textit{ConvGQR}~\cite{mo-etal-2023-convgqr} a pre-trained model for expanding the query rewrite with a potential answer; (2) \textit{T5QR}~\cite{lin2020T5QR} a T5-based query rewriting model which is trained on CANARD dataset; (3) \textit{LlaMAQR}, using \llama model as a zero-shot learner for query rewriting; (4) \textit{GPT4QR}, using \gptF model as a zero-shot learner for query rewriting; (5)\textit{\AD}, using the response generated by \ac{LLM} as a single query (inspired by \citet{gao2023-precise}); (6) \textit{HumanQR}, using the resolved-utterance by human; and (7) \textit{LLM4CS}~\cite{mao-etal-2023-large}. To ensure a fair comparison, we use the same retrieval and reranking pipeline for all baselines and our proposed methods. We reproduce the LLM4CS model, using our retrieval pipeline (i.e., GPT-4 and sparse retrieval). We use RAR with mean aggregation and with $N=5$ to reproduce LLM4CS results.\footnote{The exact experimental setup they used for reporting the results in the main table of their paper.} The prompts designed for LlaMAQR and GPT4QR models are shown in Table~\ref{tbl:prompt-qr} on page~\pageref{tbl:prompt-qr}. For the ConvGQR model, we use the code released by authors to fine-tune the model on the QReCC dataset~\cite{anantha-etal-2021-qrecc}. We use the T5QR model available on HuggingFace.\footnote{https://huggingface.co/castorini/t5-base-canard}

\begin{figure}
    \centering
    \centering
    \includegraphics[width=0.8\linewidth]{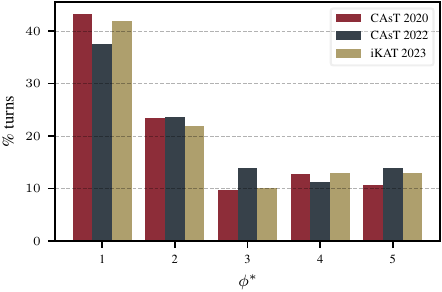}
    \caption{Distribution of the turns with the corresponding $\phi^*$ is shown. The value of $\phi^*$ is selected based on the nDCG@3 metric.}
    \label{fig:dist-phi-optimal-per-dataset}
\end{figure}

\section{Results and Discussions}
\label{sec:res}

\header{Performance comparison.}
We report the performance of our proposed models using a fixed value of $\phi=5$ (using \gptF) and the baselines in Tables~\ref{tab:retrieval-results-main-pool-all} and \ref{tab:retrieval-results-main-pool-topiocqa}. Results using \llama are provided in the Appendix (see Table~\ref{tab:retrieval-results-main-llama}, page~\pageref{tab:retrieval-results-main-llama}). 
We observe that our \QD and its variants outperform the SoTA baselines by a large margin, demonstrating the effectiveness of multi-aspect query generation. 
The better performance of  \QD compared to the LLM4CS baseline indicates (1) the importance of our proposed retrieval framework i.e., doing retrieval using each query separately and fusing the retrieved passages, and (2) the effectiveness of the generated multi-aspect queries to cover more aspects from information need of the user.
Moreover, our \QD model achieves more improvement over Recall@100 compared to nDCG@3 and Recall@10 metrics. It improves the Recall@100 compared to the best baseline 9.6\%, 6.8\%, 7.7\%, and 6.9 \% on \cast 20 \& 22, \ikat 23, and \topiocqa datasets. Hence, the generated queries cover more aspects of the information need and \QD retrieves passages from various sources of information. 
It is worth mentioning that, our proposed baseline GPT4QR is outperforming the existing baselines. The comparison between the results of GPT4QR and our proposed \QD model represents the effectiveness and importance of generating multi-aspect queries. 
The results of \AQD and \AQDAnswer are mixed over TREC datasets (see Table~\ref{tab:retrieval-results-main-pool-all}) while on \topiocqa dataset, \AQD is always better than \AQDAnswer (see Table~\ref{tab:retrieval-results-main-pool-topiocqa}).
Our \QD model has one LLM call and is more efficient than \AQD variants and is outperforming the best baselines. 

\begin{table}[]
    \caption{Human evaluation. Pairwise comparison between the queries generated by \QD and LLM4CS on TREC \ikat dialogs.}

    \centering
    \begin{tabular}{lll}
        \toprule
        & \textbf{\# turns} & \textbf{\% turns} \\ 
        \midrule
         \QD wins & 59 & 44.3\% \\
         LLM4CS wins & 37 & 27.8\% \\
         Ties & 37 & 27.8\% \\
         \bottomrule
    \end{tabular}
    \label{tab:human_annotation}
\end{table}

\header{Human evaluation.}

To further analyze whether the generated queries cover multiple aspects and how they are different from LLM4CS, we conduct human evaluation. In this study, we asked human assessors to do the side-by-side comparison of LLM4CS and MQ4CS generated queries and compare them in terms of diversity (i.e., ``Which set of queries cover more aspects of the original query?''). We asked four human annotators to assess each turn and decide the winner by the number of annotators who favored each system (e.g., 3 annotators chose MQ4CS vs.\ 1 annotator chose LLM4CS, making MQ4CS the winner). We ran the human annotation on all conversational turns of iKAT 23, and report the results in Table~\ref{tab:human_annotation}. As can be seen, in \~45\% of them MQ4CS wins the LLM4CS by generating more diverse queries, as well as \~28\% ties, confirming that our MQ4CS leads to more diverse, multi-aspect queries.

\begin{figure}
    \centering
    \includegraphics[width=0.8\linewidth]{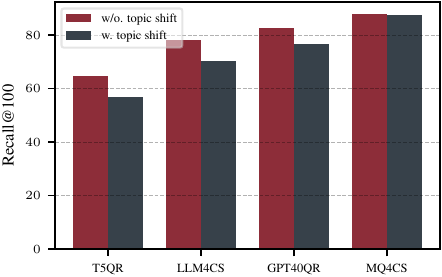}
    \caption{Performance of \QD compared with the baselines on TopiOCQA, broken into the turns with and without topic shift. }
    \label{fig:topic_shift}
\end{figure}

\header{Effect of topic shift.}
In this experiment, we particularly look at the  \topiocqa dialogues and study the performance the models on the turns with and without a topic shift, as defined in the original dataset. Figure~\ref{fig:topic_shift} shows the performance in Recall@100 of \QD and top baseline, broken by the turn type (i.e., w.\ or w/o.\ topic shift). Looking at the three baselines (T5QR, LLM4CS, and GPT4oQR), we see that they all perform better on the turns without a topic shift. These topics are presumably simpler, therefore it is not surprising to see the higher performance. Surprisingly, our proposed \QD demonstrates a different trend, where the performance of turns of the two groups is almost the same, demonstrating the power of multi-aspect query generation in tackling more complex dialog turns with topic shifts.

\begin{figure}
\centering
\begin{subfigure}{.235\textwidth}
    \centering
    \includegraphics[width=1\linewidth]{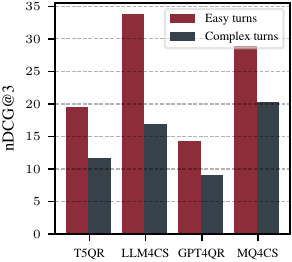}  
    \caption{\ikat 23}
    \label{fig:ikat2023}
\end{subfigure}
\begin{subfigure}{.235\textwidth}
    \centering
    \includegraphics[width=1\linewidth]{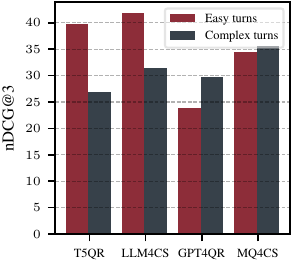}  
    \caption{\cast 22}
    \label{fig:cast2022}
\end{subfigure}
\caption{A comparison between retrieval performance of proposed \QD (with $\phi$=5) and baselines over complex and easy turns.}
\label{fig:perf_easy_dificult_phi}
\end{figure}

\header{Oracle analysis and \ourdata.}
We observe that in \AQD and \QD methods the \ac{LLM} is instructed to generate required queries and no more than $\phi$ queries, it cannot determine the optimum number of queries based on the complexity of the user utterance and is biased to generate exactly $\phi$ queries. This finding indicates the gap between query generation and document retrieval and motivates us to propose the Oracle model. The performance of our Oracle model over four conversational search datasets reported in Tables~\ref{tab:retrieval-results-main-pool-all} and \ref{tab:retrieval-results-main-pool-topiocqa} shows that using multi-aspect queries given the $\phi^*$ can lead to massive gains in the retrieval performance (up to 45\%). We release the \ourdata dataset which includes the generated queries for different values of $\phi$ and the value of $\phi^*$ over the test set of TREC \cast 19, 20, 22, \ikat 23, and \topiocqa datasets. 
To further understand the effect of multi-aspect query generation on retrieval performance, we compare the $\phi^*$ values of all the conversational turns and plot the results in Figure~\ref{fig:dist-phi-optimal-per-dataset}. In other words, we can see what percentage of the utterances are too complex to be addressed with one query rewrite. 
According to Figure~\ref{fig:dist-phi-optimal-per-dataset}, more than 55\% of \ikat and \cast 22 datasets' turns need more than one query to achieve higher nDCG@3.
The average of $\phi^*$ over turns of the \ikat 23, \cast 22, \cast 20, and \topiocqa datasets are as follows: 2.37, 2.12, 1.96, and 1.29. This finding indicates that the TREC datasets include more complex user utterances compared to the \topiocqa dataset. Although \topiocqa features various topic shifts, the TREC datasets present more complexity as they also have various topic shifts in each dialog, as well as complex information needs~\cite{dalton2020trec}. 

\header{Effect of query complexity.}
We define the utterances with $\phi^* =1$ as \textit{easy}, and $\phi^* > 1$ as \textit{complex} utterances. We report the performance of our proposed models and baselines over easy and complex turns in Figure~\ref{fig:perf_easy_dificult_phi} on \ikat 23 and \cast 22. The results on other datasets can be found in Figure~\ref{fig:perf_easy_dificult_phi_appendix} on page~\pageref{fig:perf_easy_dificult_phi_appendix}. 
As can be seen in the figure, our proposed \QD model performs better than the baseline models over complex turns. However, it is not as good as the best baseline (GPT4QR) on easy turns. This indicates that even though selecting a fixed value of $\phi$  can improve the performance over complex turns, it is not optimal for easy turns. Therefore, a per-turn $\phi$ selection model could boost the performance of \QD on easy models. We leave this extension for the future work.
In addition, according to Tables~\ref{tab:retrieval-results-main-pool-all} and \ref{tab:retrieval-results-main-pool-topiocqa}, our \QD model outperforms the best baseline model (GPT4QR) over all metrics on all datasets except nDCG@3 for \cast 20. 
This is in line with Figure~\ref{fig:dist-phi-optimal-per-dataset}, where we show 44\% of the turns of \cast 20   only require one query. So considering a constant value of $\phi$ for small datasets is sub-optimal.
Nevertheless, our Oracle model significantly outperforms the baselines and even Human-resolved utterances, reiterating the need for a flexible $\phi$ selection policy. 
For further analysis, we compare the performance of our proposed framework using $\phi$ values of 1--5 with \gptF as our \ac{LLM} and report the performance using each value of $\phi$ over \topiocqa, \cast 20 \& 22, and \ikat 23 datasets in Appendix (Tables~\ref{tab:topiocqa-ablation-phi}, \ref{tab:cast22-ablation-phi}, \ref{tab:cast20-ablation-phi}, and \ref{tab:ikat-23-ablation-phi}, respectively).
\looseness=-1

\header{Latency analysis.}
We compare the inference latency of \QD and the baselines in Appendix~\ref{app:latency} on page \pageref{app:latency}.

\section{Conclusion}
We study the effectiveness of using multi-aspect queries to enhance the retrieval for complex user queries in \ac{CIS}. 
We propose a retrieval framework that leverages an LLM (GPT-4 and \llama 3.1) to generate multi-aspect queries in the same LLM call and fuse their corresponding ranked lists.
We show the effectiveness of our proposed models over complex user utterances. Our experiments reveal that our proposed \QD framework outperforms the baselines over \cast 19, 20, 22, \ikat 23, and \topiocqa datasets. Our proposed method, while being more efficient than the SoTA LLM4CS, outperforms it significantly on all the datasets.
We show that when given the optimum value of $\phi$ for each user utterance, the Oracle model achieves massive gains, suggesting the need for designing a flexible policy for query generation. We release the generated queries and the optimal number of queries per turn to foster research in this area. In the future, we plan to leverage reinforcement-learning-based algorithms to bridge the gap between query generation and passage retrieval.

\section{Limitations}
We present to generate and use multi-aspect queries to enhance retrieval for complex user utterances. We rely on the intrinsic knowledge of \acp{LLM} to do reasoning and generate the multi-aspect queries. 
We observe that \acp{LLM} cannot determine the efficient number of multi-aspect queries for each user utterance based on its complexity. This is because of the gap between retrieval and query generation. We did not study other biases of different \acp{LLM} on query generation and response generation.
When generating multi-aspect queries per utterance (up to $\phi=5$), latency can increase by a factor of $\phi$. However, the retrieval pipeline can be done in parallel, without extra cost compared to one query rewrite. 
Further details about the latency of our pipeline can be found in Appendix~\ref{app:latency} on page \pageref{app:latency}.

\section{Ethical considerations}

Stressing the need to study and measure biases in Language Models (LLMs) when generating data, we think it could cause unexpected ethical issues.
Consequently, we need to study the potential biases that exist in the data and formalize their impact on the final output of the model.
While in this study we propose to use the answers and queries generated by LLMs for retrieval models, we think these methods should be used carefully in real-world retrieval systems, and designers should consider these biases.

\bibliography{main}

\clearpage
\appendix
\textbf{Appendix}

\section{Experimental Setup}
\label{sec:appendix-exp-setup}
    
\begin{table}[]
\caption{Statistics of the datasets.}

\small
    \centering
    \adjustbox{max width=0.4\textwidth}{
    \begin{tabular}{lp{1cm}cp{1cm}}
\toprule    
\textbf{Dataset} & \textbf{\#Conv.} & \textbf{\#Turns} & \textbf{Conv.\ Length} \\
\toprule
\cast 20  & \phantom{0}25 & \phantom{0,}216 & \phantom{0}8.6 \\
\cast 22  & \phantom{0}18 & \phantom{0,}205 & 11.39 \\
\ikat 23  & \phantom{0}25 & \phantom{0,}133 & 13.04 \\
\topiocqa   & 205 & 2,514 & 12 \\
\bottomrule
       
    \end{tabular}
    }
    \label{tab:dataset-stat}
\end{table}

    \header{{Hyper-parameters.}} For the first-stage retrieval we employ the BM25 model from Pyserini \cite{Lin_etal_SIGIR2021_Pyserini} using the default values for the parameters. For the reranker, we use the pre-trained Cross-Encoder model \texttt{ms-marco-MiniLM-L-6-v2} from the \texttt{sentence\_transformers} library with a maximum length of 512. In \AQD and \QD approaches we take the top 1000 passages returned by BM25 and pass them to reranker. After interleaving the results, we take the top 1000 passages. In \AQDAnswer method, we interleave the top passages returned by BM25 and pass the first 1000 passages to reranker. In other \QR and \AD baselines, we rerank the top 1000 passages returned by BM25.
    We use \texttt{Llama-3.1-8B-Instruct} with the following parameters: \texttt{top\_k=$10$}, \texttt{top\_p=$0.9$}, \texttt{temperature=$0.75$}. We conduct our experiments on a single A6000 GPU with 32 GB RAM.  
    We use the \gptF model as a zero-shot learner using the default values of parameters for all approaches. We use \gptF-O model for \topiocqa dataset. 
    
    \header{{Dataset.}}  We report the results on \topiocqa, TREC \ikat 23, TREC \cast 20 and 22 datasets. The statistics of these datasets are shown in Table~\ref{tab:dataset-stat}. The \topiocqa is a large-size open-domain conversational dataset that incorporates topic shifts and is based on Wikipedia \cite{adlakha-etal-2022-topiocqa}.
    The TREC \ikat 23 dataset is one of the few datasets that features complex dialogues where single-query rewriting is not effective. The average length of conversations in \ikat is 13.04 which makes the context modeling task more challenging.  \looseness=-1

    \header{Metrics.} We evaluate passage retrieval performance using the official metrics used in the literature, namely, nDCG@3, Recall@100, and MRR. nDCG@3 evaluates the scenarios where the top passages are intended to be presented to the user. We calculate these metrics using the {\fontfamily{qcr}\selectfont{trec\_eval}} tool. We do the statistical significance tests using paired t-tests at $p<0.05$ level and compare the best model with other models and baselines.

\section{Experimental results on TREC CAsT 19}
In this section, we report the results of TREC CAsT 19. We include these results in the appendix for space considerations. Moreover, given the relative simplicity of TREC CAsT 19 compared to other CAsT collections and iKAT, we find these results non-centric to our paper. Table~\ref{tab:retrieval-results-main-pool-cast19} reports the results of the baselines, as well as our proposed methods in terms of our main evaluation metrics. As we see in the table, the superiority of rewriting techniques already shows the simplicity of the CAsT 19 collection, where the dialogues are not long enough and user utterances are only dependent on their own previous utterance (as opposed to system responses). We see that our proposed method perform  

\begin{table}[t]
\caption{Passage retrieval results on the TREC CAsT 19 dataset with $\phi=5$. The best results that are significantly better (t-tests with $p\_value \leq 0.05$) are in \textbf{bold} face.}
\centering

\adjustbox{max width=0.5\textwidth}{
 \begin{tabular}{lllll}
\toprule
\textbf{Method} & \textbf{nDCG@3} & \textbf{R@100} & \textbf{MRR} \\
\toprule

GPT4QR     &  \textbf{56.5}  & 62.0  & 73.8     \\
T5QR       &  \textbf{56.5}  & 59.8  & \textbf{74.6}    \\
\ConvGQR   &  50.2  & 53.6  & 67.5       \\
GPT4-\AD   &  47.9  & 53.9  & 68.8  \\
LLM4CS     &  52.2 & 55.8 & 72.7 \\
\midrule
\AQD  &  48.6 & 60.0 & 74.2  \\  
\AQDAnswer & 48.0 & 54.8 & 68.8    \\
\QD   & 49.6 & \textbf{62.2} & 73.9   \\
\midrule
 \multicolumn{5}{c}{\textbf{Oracle}} \\
\midrule
\AQD  & 68.7 & 68.9 & 88.5   \\
\AQDAnswer &  52.3 & 61.3 & 73.5 \\
\QD       & 67.8 & 70.8 & 84.4   \\

\bottomrule

\end{tabular}
}
\label{tab:retrieval-results-main-pool-cast19}

\end{table}

\section{Experiments on our framework using different values of $\phi$}
\label{sec:app-different-phi}
\begin{table}[t]

\caption{Passage retrieval results of our \QD framework on \ikat 23 using different values of $\phi$ and \gptF model as \ac{LLM}.}
\centering
\adjustbox{max width=0.5\textwidth}{
 \begin{tabular}{lllllll}
\toprule
\textbf{} & $\phi$ & \textbf{nDCG@3} & \textbf{nDCG} & \textbf{R@10} & \textbf{R@100} & \textbf{MRR}  \\
\toprule

\multirow{5}{*}{\rotatebox[origin=c]{90}{\AQD}} & 1  & 25.1  & 26.3  & 6.6   & 21.6  & 47.0   \\
 & 2  & 24.9  & 30.0  & 8.0   & 24.6  & 50.1   \\
 & 3  & 24.8  & 31.7  & 7.2   & 24.9  & 51.8   \\
 & 4  & 23.6  & 31.1  & 7.2   & 24.3  & 49.8   \\
 & 5  & 23.0  & 29.9  & 6.8   & 22.5  & 50.3   \\

\midrule

\multirow{5}{*}{\rotatebox[origin=c]{90}{\parbox{1.8cm}{\AQD \\+rerank}}} & 1 & 16.1  & 20.6  & 5.3   & 20.8  & 37.1   \\
 & 2 & 18.6  & 24.9  & 6.2   & 23.3  & 43.4   \\
 & 3 & 18.4  & 26.9  & 5.9   & 23.4  & 39.8   \\
 & 4 & 17.3  & 27.4  & 5.5   & 22.8  & 36.2   \\
 & 5 & 18.1  & 26.8  & 4.9   & 21.2  & 42.9   \\

\midrule

\multirow{5}{*}{\rotatebox[origin=c]{90}{\QD}} & 1   & 28.2  & 28.8  & 6.9   & 22.8  & 52.0   \\
 & 2   & 25.3  & 29.4  & 6.7   & 22.7  & 52.7   \\
 & 3   & 22.0  & 28.9  & 6.6   & 22.6  & 50.2   \\
 & 4   & 24.2  & 30.1  & 6.8   & 23.3  & 53.0   \\
 & 5   & 22.6  & 31.1  & 6.4   & 25.1  & 52.8   \\

\bottomrule
\end{tabular}

}
\label{tab:ikat-23-ablation-phi}
\end{table}


\begin{table}[t]
\caption{Passage retrieval results of our \QD framework on \cast 22 using different values of $\phi$ and \gptF model as \ac{LLM}.}

\centering
\adjustbox{max width=0.5\textwidth}{
 \begin{tabular}{lllllll}
\toprule
\textbf{} & $\phi$ & \textbf{nDCG@3} & \textbf{nDCG} & \textbf{R@10} & \textbf{R@100} & \textbf{MRR}  \\
\toprule

\multirow{5}{*}{\rotatebox[origin=c]{90}{\AQD}} & 1  & 39.2  & 37.1  & 8.2   & 29.8  & 68.5   \\
 & 2  & 38.0  & 40.2  & 7.5   & 30.2  & 68.5   \\
 & 3  & 36.1  & 42.1  & 7.9   & 31.3  & 69.2   \\
 & 4  & 35.1  & 42.6  & 7.3   & 31.6  & 67.4   \\
 & 5  & 33.2  & 42.3  & 7.0   & 31.1  & 68.7   \\

\midrule

\multirow{5}{*}{\rotatebox[origin=c]{90}{\parbox{1.8cm}{\AQD \\+rerank}}} & 1 & 35.6  & 35.5  & 7.6   & 27.2  & 66.6   \\
 & 2 & 32.3  & 39.5  & 7.4   & 29.1  & 61.5   \\
 & 3 & 32.8  & 41.6  & 7.6   & 29.3  & 61.3   \\
 & 4 & 32.7  & 43.4  & 7.5   & 29.7  & 62.6   \\
 & 5 & 32.5  & 41.8  & 7.4   & 31.1  & 60.5   \\

\midrule

\multirow{5}{*}{\rotatebox[origin=c]{90}{\QD}} & 1   & 41.0  & 38.0  & 8.5   & 30.6  & 71.1   \\
 & 2   & 40.4  & 42.3  & 8.2   & 32.3  & 75.0   \\
 & 3   & 35.9  & 43.2  & 7.7   & 32.4  & 70.6   \\
 & 4   & 34.9  & 43.7  & 7.8   & 32.4  & 67.2   \\
 & 5   & 35.0  & 45.1  & 8.0   & 32.5  & 71.8   \\

\bottomrule
\end{tabular}

}
\label{tab:cast22-ablation-phi}

\end{table}


\begin{table}[t]
\caption{Passage retrieval results of our \QD framework on \cast 20 using different values of $\phi$ and \gptF model as \ac{LLM}.}

\centering
\adjustbox{max width=0.5\textwidth}{
 \begin{tabular}{lllllll}
\toprule
\textbf{} & $\phi$ & \textbf{nDCG@3} & \textbf{nDCG} & \textbf{R@10} & \textbf{R@100} & \textbf{MRR}  \\
\toprule

\multirow{5}{*}{\rotatebox[origin=c]{90}{\AQD}} & 1  & 45.9  & 53.8  & 19.1  & 55.7  & 72.2   \\
 & 2  & 47.6  & 58.0  & 20.0  & 58.4  & 76.8   \\
 & 3  & 43.6  & 58.5  & 19.7  & 59.5  & 75.3   \\
 & 4  & 42.0  & 58.2  & 19.3  & 60.0  & 72.4   \\
 & 5  & 44.8  & 59.6  & 19.5  & 60.9  & 77.4   \\
\midrule

\multirow{5}{*}{\rotatebox[origin=c]{90}{\parbox{1.8cm}{\AQD \\+rerank}}}  & 1 & 45.9  & 52.4  & 18.6  & 50.4  & 74.0   \\
 & 2 & 44.4  & 54.9  & 17.7  & 51.3  & 71.4   \\
 & 3 & 42.3  & 55.3  & 17.9  & 50.6  & 71.8   \\
 & 4 & 42.8  & 56.3  & 18.0  & 50.6  & 67.6   \\
 & 5 & 45.0  & 57.6  & 18.2  & 54.5  & 73.3   \\
\midrule

\multirow{5}{*}{\rotatebox[origin=c]{90}{\QD}} & 1   & 47.5  & 57.2  & 20.8  & 60.9  & 75.8   \\
 & 2   & 47.0  & 59.9  & 21.1  & 63.5  & 77.4   \\
 & 3   & 45.0  & 60.2  & 20.6  & 63.1  & 76.5   \\
 & 4   & 43.3  & 59.1  & 19.3  & 62.7  & 76.8   \\
 & 5   & 44.8  & 60.5  & 20.6  & 62.5  & 77.7   \\

\bottomrule
\end{tabular}
}
\label{tab:cast20-ablation-phi}

\end{table}


\begin{table}[t]
\caption{Passage retrieval results of our \QD framework on \topiocqa using different values of $\phi$ and \gptF model as \ac{LLM}.}

\centering
\adjustbox{max width=0.38\textwidth}{
 \begin{tabular}{llllll}
\toprule
\textbf{} & $\phi$ & \textbf{mAP}  &  \textbf{R@10}  &  \textbf{R@100}  &  \textbf{MRR}   \\
\toprule

\multirow{5}{*}{\rotatebox[origin=c]{90}{\AQD}} & 1  &    46.5  & 68.2  & 81.7  & 46.5  \\
 & 2  &   47.2  & 69.8  & 85.2  & 47.2 \\
 & 3  &  46.7  & 70.2  & 86.0  & 46.7  \\
 & 4  &  46.5  & 70.3  & 86.6  & 46.5  \\
 & 5  & 46.7  & 70.6  & 87.0  & 46.7 \\
\midrule

\multirow{5}{*}{\rotatebox[origin=c]{90}{\parbox{1.8cm}{\AQD \\+rerank}}}  & 1 & 43.3  & 64.6  & 80.8  & 43.3  \\
 & 2  &  43.6  & 65.5  & 83.8  & 43.6  \\
 & 3  & 43.4  & 65.2  & 83.9  & 43.4   \\
 & 4  &  43.4  & 65.1  & 83.9  & 43.4   \\
 & 5  &  43.7  & 65.3  & 84.4  & 43.7\\
\midrule

\multirow{5}{*}{\rotatebox[origin=c]{90}{\QD}} & 1   &  47.1  & 67.9  & 82.9  & 47.1 \\
 & 2  &  48.0  & 70.8  & 85.0  & 48.0  \\
 & 3  &  48.4  & 72.2  & 87.0  & 48.4  \\
 & 4  &  47.4  & 72.2  & 86.9  & 47.4  \\
 & 5  & 47.5  & 72.6  & 87.8  & 47.5 \\

\bottomrule
\end{tabular}
}
\label{tab:topiocqa-ablation-phi}

\end{table}

\begin{figure*}
\centering
\begin{subfigure}{.45\textwidth}
    \centering
    \includegraphics[width=.95\linewidth]{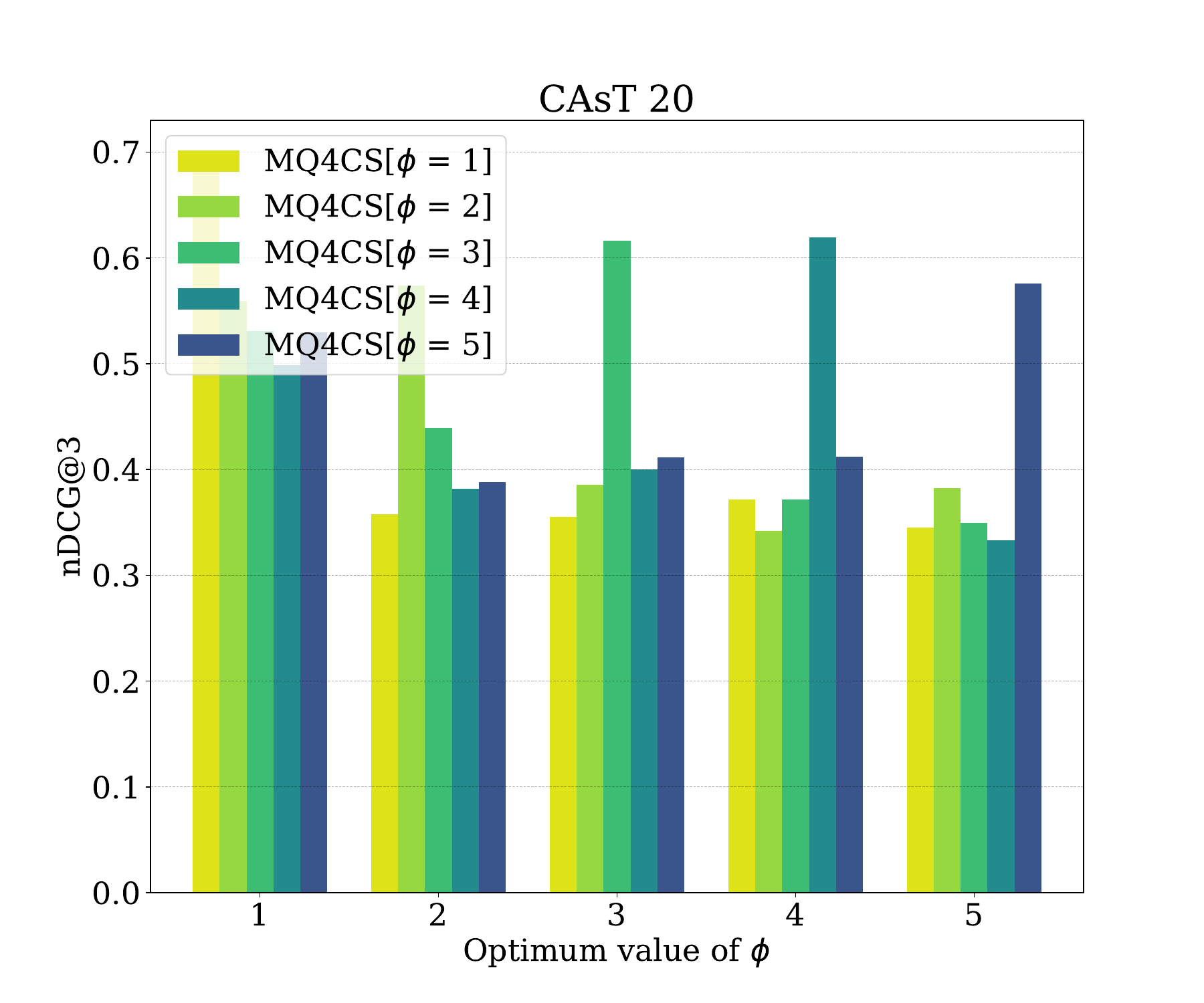}  
    \caption{\cast 20}
    \label{fig:cast20}
\end{subfigure}
\begin{subfigure}{.45\textwidth}
    \centering
    \includegraphics[width=.95\linewidth]{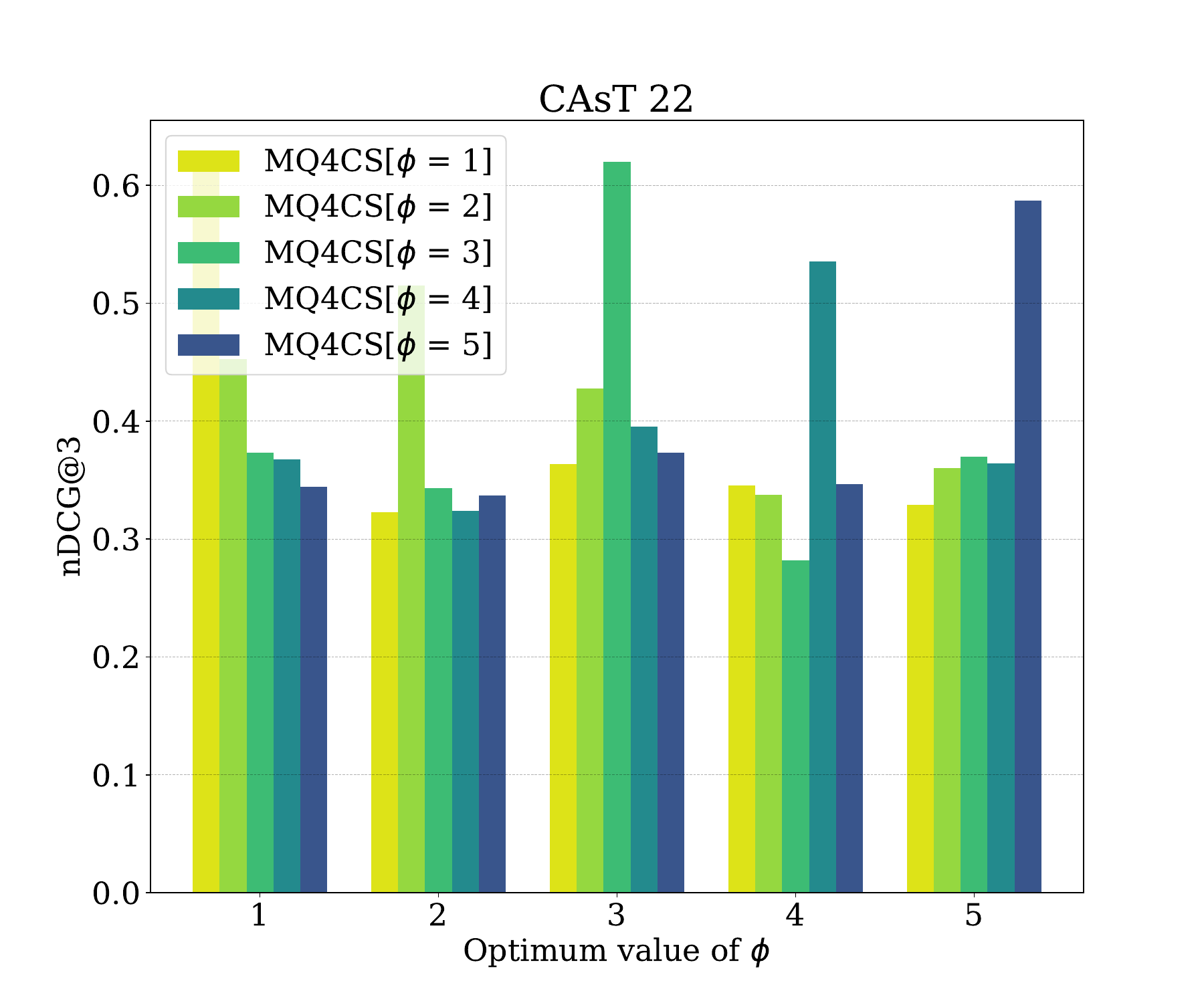}  
    \caption{\cast 22}
    \label{fig:cast22}
\end{subfigure}
\begin{subfigure}{.45\textwidth}
    \centering
    \includegraphics[width=.95\linewidth]{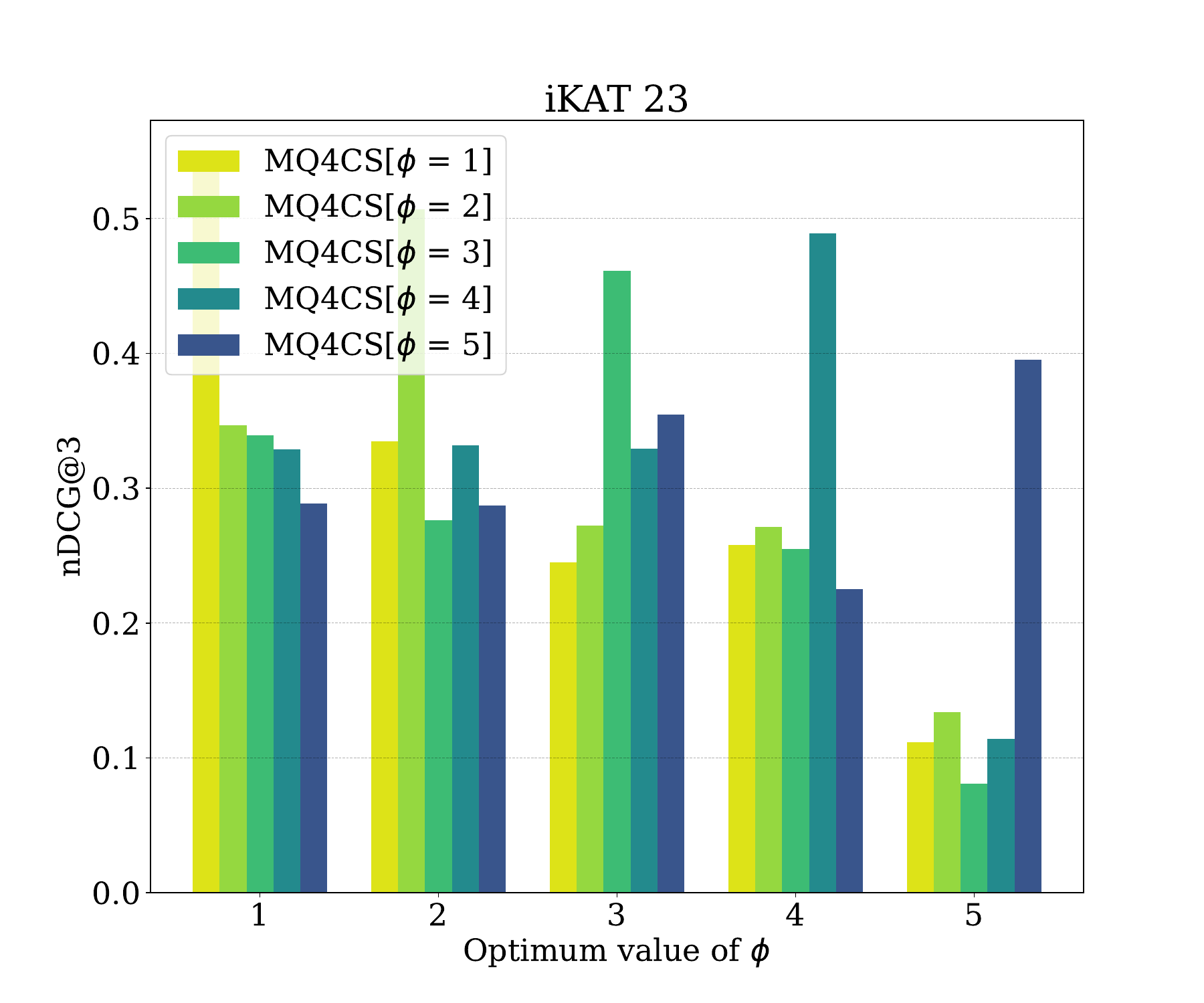}  
    \caption{\ikat 23}
    \label{fig:ikat23}
\end{subfigure}
\begin{subfigure}{.45\textwidth}
    \centering
    \includegraphics[width=.95\linewidth]{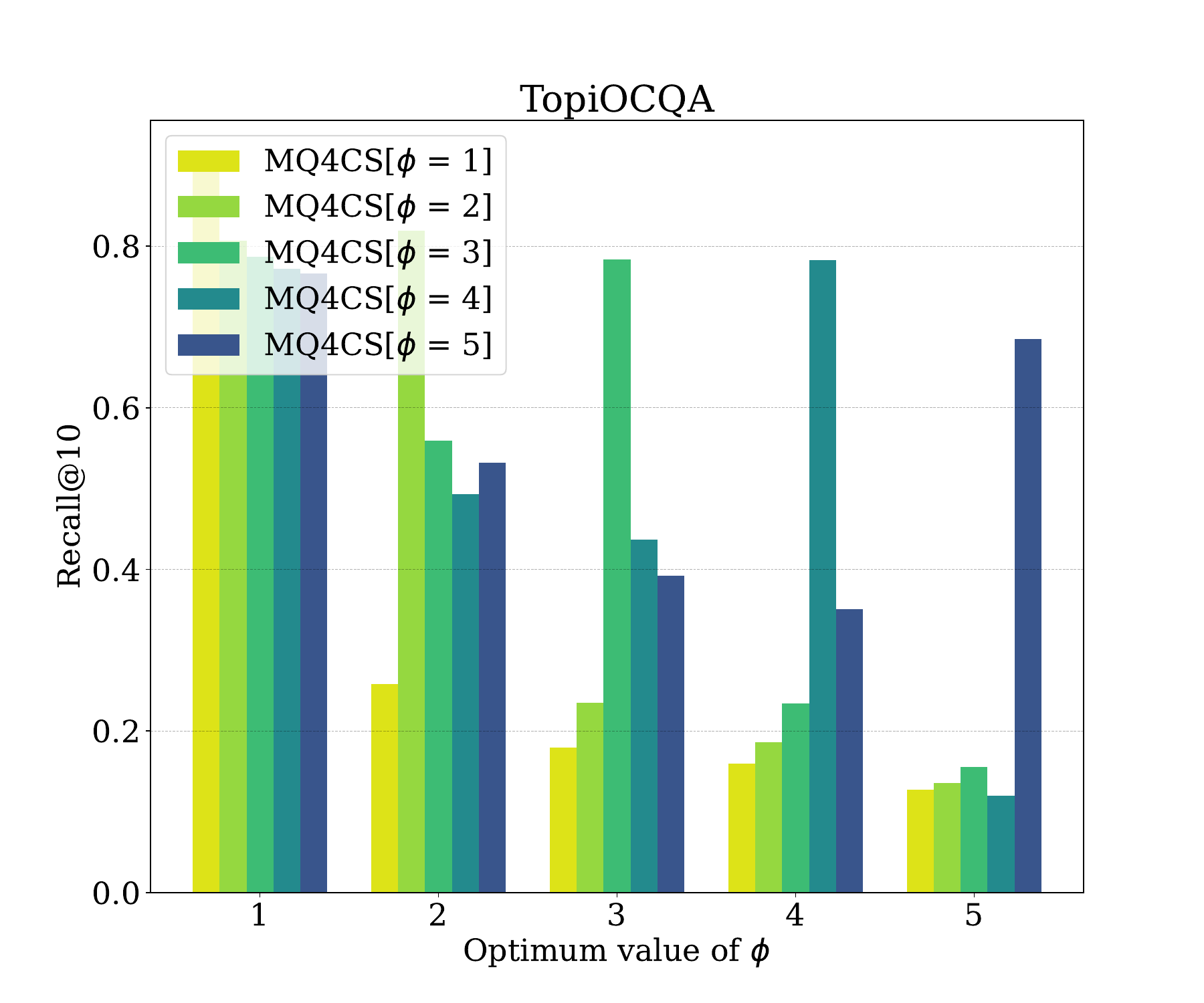}  
    \caption{\topiocqa}
    \label{fig:topiocqa}
\end{subfigure}
\caption{Performance of \QD model using different constant values of $\phi$ per turns with different optimum values of $\phi$.}
\label{fig:perf_diff_phi}
\end{figure*}

We report the performance of our proposed models including \AQD, \AQDAnswer, and \QD using different values for $\phi$ parameter in Tables \ref{tab:cast20-ablation-phi}, \ref{tab:cast22-ablation-phi}, \ref{tab:ikat-23-ablation-phi} and \ref{tab:topiocqa-ablation-phi}. In these experiments, we keep the $\phi$ value consistent for all conversational turns (i.e., user utterance). As can be seen, we have consistent results using different constant values of $\phi$.
We show the performance of the \QD model using each different constant value of $\phi$ over optimum values of $\phi$ based on the Oracle model in Figure \ref{fig:perf_diff_phi}. As can be seen in Figure \ref{fig:perf_diff_phi}, using a constant value of $\phi$, we achieve the best performance over the turns with the same optimum value of $\phi$. This finding justifies the similar performance of the \QD model when using different constant values for $\phi$.

\section{Experiments on our framework
using different LLMs}
\label{sec:different-llm}

\begin{table*}[t]
\caption{Experiments with Llama3.1 as baseline LLM in our models on TREC \cast 20 and 22 and \ikat 23 datasets. In these experiments, we use the $\phi=5$.}

\centering
\adjustbox{max width=1\textwidth}{
 \begin{tabular}{llllllllllll}
\toprule
 \multirow{2}{*}{\textbf{Method}} & \multicolumn{3}{c}{\textbf{\cast 20}} & & \multicolumn{3}{c}{\textbf{\cast 22}} & & \multicolumn{3}{c}{\textbf{\ikat 23}}\\  \cmidrule{2-4} \cmidrule{6-8} \cmidrule{10-12}

& \textbf{nDCG@3} & \textbf{R@100} & \textbf{MRR} & & \textbf{nDCG@3}  & \textbf{R@100} & \textbf{MRR} & & \textbf{nDCG@3}  & \textbf{R@100} & \textbf{MRR}  \\
\cmidrule(lr){1-12} 
LLaMAQR  & 36.3  & 49.3  & 51.4    &&  30.8  & 27.3  & 49.0   &&  9.6   & 14.2  & 16.0\\
\llama-\AD   & 19.3  & 31.0  & 29.8    && 21.1  & 21.5  & 34.4    && 10.6  & 18.8  & 16.9 \\
\cmidrule{1-12}
HumanQR    & 50.5   & 61.8  & 66.8   && 41.3   & 37.6  & 60.4   && 30.7   & 35.8  & 43.3 \\
\cmidrule{1-12}
\cmidrule{1-12}
\llama-\AQD  & 28.4  & 48.1  & 48.2            && 27.4  & 28.6  & 48.2    && 12.3  & 14.9  & 22.0 \\
\llama-\AQDAnswer &22.9  & 33.3  & 35.1      && 34.6  & 11.1  & 55.1    && 11.3  & 16.2  & 18.2\\
\llama-\QD   & 36.0  & 57.7  & 57.0    && 30.4  & 34.7  & 54.0    && 17.0  & 16.2  & 30.8 \\


\bottomrule
\end{tabular}
}
\label{tab:retrieval-results-main-llama}

\end{table*}

We report additional results here using \llama3.1 as base LLM for multiple query generation. Overall the trends and findings are similar than with the base \gptF model.
More specifically, we observe that our \QD with \llama can outperform the comparable single query baselines over several metrics. 
Also, we observe that \QD performs better than \AQD on all metrics on TREC datasets, similarly to \gptF.

It is worth mentioning here that the \llama3.1 model is instructed given few-shot prompts and is not fine-tuned for this task.
The few-shot prompts designed for \llama model in \AQD and \QD approaches are shown in Tables~\ref{tbl:prompt-aqd-llama} and \ref{tbl:prompt-qd-llama}, respectively.

\section{Analysis of the query complexity}
In this section, we include the results of this analysis on \cast 20 and \topiocqa in addition to the results already presented in Table~\ref{fig:perf_easy_dificult_phi} on page~\pageref{fig:perf_easy_dificult_phi}. On these two datasets, we observe a similar pattern, where our \QD outperforms the SoTA baselines when addressing complex queries. We still see room for improving the simpler queries which calls for having a flexible $\phi$ selection policy.

\begin{figure}
\centering
\begin{subfigure}{.235\textwidth}
    \centering
    \includegraphics[width=1\linewidth]{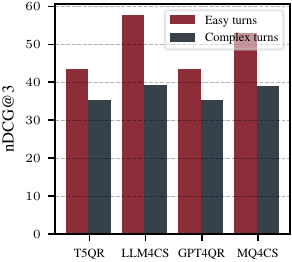}  
    \caption{\cast 20}
    \label{fig:cast2020}
\end{subfigure}
\begin{subfigure}{.235\textwidth}
    \centering
    \includegraphics[width=1\linewidth]{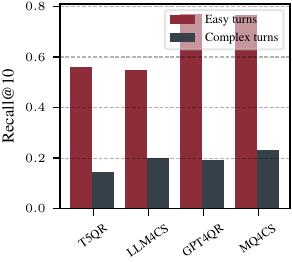}  
    \caption{\topiocqa}
    \label{fig:topicqa}
\end{subfigure}
\caption{A comparison between retrieval performance of proposed \QD (with $\phi$=5), and baselines over complex and easy turns. on \cast 20 and \topiocqa.}
\label{fig:perf_easy_dificult_phi_appendix}
\end{figure}

\section{Prompts}
\label{sec:appendix}
The prompt used for \AQD and \AD approach using \gptF is shown in Table \ref{tbl:prompt-aqd-ad-gpt4}. We use the same prompt for \AQDAnswer approach.
The prompt used for zero-shot \QD using \gptF model is shown in Table \ref{tbl:prompt-qd-gpt4}. The term $ctx$ in the prompts designed for \gptF includes all of the previous user utterances and system responses. 

For \QR model the prompt shown in Table \ref{tbl:prompt-qr} is designed. This prompt is used for both \llama and \gptF models. We pass all the previous user and system interactions as $ctx$ in this prompt.

The two-shot prompt designed for \llama \AQD approach is shown in Table \ref{tbl:prompt-aqd-llama}. In this prompt, the answer generated by \llama itself is passed as the $response$.
For the prompts of \llama model, all the previous user utterances with the last system response are passed as context to the model. 

The one-shot prompt designed for \llama \QD approach is shown in Table \ref{tbl:prompt-qd-llama}.

\section{Latency}
\label{app:latency}

For query generation, We use the OpenAI API and we do not have access to the latency of the model. But we are issuing only one prompt for \OurModel model and 2 prompts for \AQD and \AQDAnswer.

The latency of our \OurModel model for query generation using LLaMA (3.1 with 8B parameters) as our LLM is on average 7.581(s) while the latency of LLM4CS model for the similar approach (RAR) is 11.84 (s) using LLaMA. Even though our latency is this much but it’s on par (or better) than existing methods such as LLM4CS.

For \AQD and \AQDAnswer
 models our query generation latency is, 16,592 s (7,252 + 9,34) on average because we call LLM one time for answer generation and then for query generation using the answer. Which is still better than the latency of LLM4CS on RTR approach.

In addition, we provide the retrieval latency of \OurModel in Table~\ref{tab:app:latency}. To report retrieval latency, we try to simulate the real-world scenario. This means we run the whole pipeline using one user utterance. We repeat this process for 20 different user utterances. We select the user utterances from different topics. We report in Table~\ref{tab:app:latency} the average run time for the selected 20 user utterances.

As can be seen, the total retrieval latency of our \OurModel model is comparable to the latency of T5QR model. However, \OurModel uses 5 queries for retrieval while T5QR uses a single query, the latency of \OurModel is much lower than T5QR. We justify it as follows: 1) The queries generated by \OurModel are shorter than query rewrites generated by other baseline models (including T5QR), this is important because BM25 is sensitive to the number of unique query words. We refer the reviewer to the provided dataset. 2) The phi queries generated by \OurModel share common words and as we run all of them in one batch, the model benefits from caching. which means for repetitive words, the model does not have any additional computational costs.

The aforementioned results are obtained using the following hardware: 32 CPUs : Intel(R) Xeon(R) Gold 5118 CPU @ 2.30GHz 1 nvidia\_rtx\_a6000 with 32 gb memory

\begin{table}[t]
\caption{Latency of our framework in seconds, compared to baselines.}

\centering
\adjustbox{max width=0.49\textwidth}{
\begin{tabular}{lccc}
\toprule
\textbf{Model} & \textbf{Retrieval} & \textbf{Re-ranking} & \textbf{Total} \\
\midrule
T5QR & 14.250 & \phantom{0}1.819 & 16,069 \\
$\text{MQ4CS}$ & \phantom{0}8.560 & \phantom{0}8.930 & 17,490 \\
$\text{MQ4CS}_{ans}$ & 14.223 & \phantom{0}8.698 & 22,921 \\
$\text{MQ4CS}_{ans} + \text{rerank}$ & 14.223 & 10,639 & 24,862 \\
\bottomrule
\end{tabular}
}
\label{tab:app:latency}
\end{table}

\begin{table}[!h]
        \vspace{1em}
        \caption{The prompt designed for \AQD and \AD approaches using \gptF as a zero-shot learner.}
        \begin{tabularx}{\linewidth}{X}
            \toprule
            \bf (1) Initial Answer Generation and (2) Query Generation in \AQD approach. \\
            \midrule
            (1) \# Instruction:\\\textit{I will give you a conversation between a user and a system. Also, I will give you some background information about the user. You should answer the last question of the user. Please remember that your answer to the last question of the user shouldn't be more than 200 words.}\\\\
            
            \# Background knowledge: \{$ptkb$\}\\ 
            \# Context: \{$ctx$\}\\ 
            \# User question: \{$user\ utterance $\} \\ 

            \# Response: \\ \\

            \midrule
            
            (2) \# Can you generate the unique queries that can be used for retrieving your previous answer to the user? (Please write each query in one line and don't generate more than $\phi$ queries)\\ \\

            \# Generated queries:\\ \\ 
            \bottomrule
        \end{tabularx}
        \label{tbl:prompt-aqd-ad-gpt4}
\end{table}

\begin{table}[!h]
        \vspace{1em}
\caption{The prompt designed for \QR using \gptF and \llama models as a zero-shot learner.}

        \begin{tabularx}{\linewidth}{X}
            \toprule
            \bf Query re-writing (\QR).\\
            \midrule
            \# Instruction:\\\textit{I will give you a conversation between a user and a system. Also, I will give you some background information about the user. You should rewrite the last question of the user into a self-contained query.}\\\\
            \# Background knowledge: \{$ptkb$\}\\ 
            \# Context: \{$ctx$\}\\
            \# Please rewrite the following user question: \{$user\ utterance $\} \\ 

            \# Re-written query: \\ \\
            
            \bottomrule
        \end{tabularx}
        \label{tbl:prompt-qr}
\end{table}

\begin{table}[!h]
        \vspace{1em}
                \caption{The prompt designed for \QD approach using \gptF as a zero-shot learner.}
        \begin{tabularx}{\linewidth}{X}
            \toprule
            \bf Query Generation in \QD approach. \\
            \midrule
            \# Instruction:\\\textit{I will give you a conversation between a user and a system and some background information about the user. Imagine you want to find the answer to the last user question by searching Google. You should generate the unique search queries that you need to search in Google. Please don't generate more than $\phi$ queries and write each query in one line.
}\\\\
            \# Background knowledge: \{$ptkb$\}\\ 

            \# Context: \{$ctx$\}\\ 
            \# User question: \{$user\ utterance $\} \\ \\

            \# Generated queries: \\ \\
            
            \bottomrule
        \end{tabularx}

        \label{tbl:prompt-qd-gpt4}
\end{table}

\begin{table*}[!h]
    \small        
        \vspace{1em}
                \caption{The prompt designed for two-shot \AQD using \llama model.}
        \begin{tabularx}{\linewidth}{X}
        
            \toprule
            \bf The two-shot prompt for query generation in \AQD. \\
            \midrule
            \# Instruction:\\\textit{Generate the unique queries to search them in a search engine to retrieving the last response of the system to the user. (Please write each query in one line and don't generate more than $\phi$ queries)}\\\\
            \#  Example 1\\
            
            \# Background knowledge: 1: My sister is following the `West Worl', but I don't like it, 2: Johnny Depp made the Pirates of the Caribbean excellent, 3: My friend suggested to me the `Now you see me' movie, ...\\ 
           
            \# Context: \\
            \hspace{10pt}user: Can you tell me what the Golden Globe Awards is? \\
            \hspace{10pt}user: What is it?\\
            \hspace{10pt}user: Is it different from the Oscars?\\
            \hspace{10pt}user: What is the difference between them?\\
            \hspace{10pt}user: No, I mean Academy Awards and Golden Globe Awards.\\
            \hspace{10pt}user: What else?\\
            \hspace{10pt}system: The Hollywood Foreign Press Association, a group of 93 journalists from around 55 countries, ....  \\
            \# User question:  Did any of my favorite actresses win any of them?  \\ 
            \# System response: Yes, both Jennifer Aniston and Lisa Kudrow, who you enjoyed in the Friends series, have won Golden Globe Awards. Jennifer Aniston won the Golden Globe in 2003 for Best Actress in a Television Series – Musical or Comedy for Friends. Lisa Kudrow also won this category in 1998 for the same series. Aniston's performance in The Morning Show also earned her another nomination in 2020. On the other hand, neither of the actresses have won an Academy Award as of yet.  \\ 
            \# Generated queries: \\
            \hspace{10pt}1. Has Jennifer Aniston ever won a Golden Globe or an Academy Award? \\
            \hspace{10pt}2. Has Lisa Kudrow ever won a Golden Globe or an Academy Award?\\ 
            \hspace{10pt}3. List of Golden Globe winners in Best Actress in a Television Series – Musical or Comedy category for 1998 and 2003.\\
            \hspace{10pt}4. Did Jennifer Aniston win any awards for `The Morning Show'? \\
 \\ \hdashline \\

            \#  Example 2\\
            \# Background knowledge: 1: I am diagnosed with diabetes type 2, 2: My husband is a light drinker, 3: We have pasta twice a week! It is my favorite dish,.... \\ 
            
            \# Context: \\
            \hspace{10pt}user: Can you tell me about different types of alcoholic drinks?\\
            \hspace{10pt}user: What is the difference between them?\\
            \hspace{10pt}user: No, the other category.\\
            \hspace{10pt}user: Ok. Can you compare base liquors and the third one?\\
            \hspace{10pt}user: Interesting, but I was mainly looking for ingredients and flavor! FYI, I don't drink myself.\\
            \hspace{10pt}user: How is it different from Liqueurs?\\
            \hspace{10pt}user: Why are their names so similar? \\
            \hspace{10pt}user: How about the percentage of alcohol? \\
            \hspace{10pt}system: Liquor is hard (the hardest) alcohol product made by distillation, often clocking ....  \\ 

            \# User question: Which types are suitable for my husband? \\ 
            \# System response: Considering your husband is a light drinker, he might enjoy lower-alcohol content beverages such as wine, beer, or certain liqueurs. Specifically, since he enjoys seafood like salmon, white wine such as chardonnay, pinot grigio, or a light-bodied beer might be a good match. However, always remember to consume alcohol in moderation, and if there are any health concerns, consult with a physician.  \\ 
            \# Generated queries:  \\
            \hspace{10pt}1. Which alcoholic beverages are suitable for a light drinker who likes salmon? \\
            \hspace{10pt}2. What types of alcohol go well with seafood? \\
            \hspace{10pt}3. What types of alcohol are preferred for people who drink lightly? \\
            \hspace{10pt}4. What alcoholic drinks have lower alcohol content? \\
            \hspace{10pt}5. Recommendations for alcoholic beverages for light drinkers.  \\\\

            \hdashline \\
            \#  Example 3\\
            \# Background knowledge: \{$ptkb$\}\\ 
            \# Context: \{$ctx$\}\\
            \# User question: \{$user\ utterance $\} \\
            \# System response: \{$response$\} \\  
            \# Generated queries:  \\
            
            \bottomrule
        \end{tabularx}

        \label{tbl:prompt-aqd-llama}
\end{table*}

\begin{table*}[!h]
 \small    
        \vspace{1em}
                \caption{The prompt designed for one-shot \QD using \llama model.}
        \begin{tabularx}{\linewidth}{X}
            \toprule
            \bf The one-shot prompt designed for \QD. \\
            \midrule
            \# Instruction:\\\textit{Please generate self-contained unique questions that should be searched in a search engine to answer the user's LAST utterance. (Please write each query in one line and don’t generate more than $\phi$ queries)}\\\\
            \# Example 1\\
            \# Background knowledge: 1: My sister is following the `West Worl', but I don't like it, 2: Johnny Depp made the Pirates of the Caribbean excellent, 3: My friend suggested to me the `Now you see me' movie, it was fantastic, 4: I went on a biking trip last year, 5: I usually like to drink coffee in the morning, 6: I watched the proposal and enjoyed it. Ryan Reynolds is my favorite!, 7: The `Friends' series was terrific, Jennifer Aniston and Lisa Kudrow were the best stars!\\ \\
            \# Context: \\
                \hspace{10pt}user: Can you tell me what the Golden Globe Awards is? \\
                 \hspace{10pt}user: What is it? \\
                 \hspace{10pt}user: Is it different from the Oscars? \\
                 \hspace{10pt}user: What is the difference between them?\\
                 \hspace{10pt}user: No, I mean Academy Awards and Golden Globe Awards. \\
                 \hspace{10pt}user: What else?\\
                  \hspace{10pt}system: The Hollywood Foreign Press Association, a group of 93 journalists from around 55 countries, are the committee for the Globes. On the contrary, the voting body of the Academy Awards; the Academy of Motion Picture Arts and Sciences (AMPAS) consists of 6,000 voting members. The Golden Globes Award recognizes the excellence of artists in both the film and television industry in the United States as well as in other countries. However, Academy Awards only recognize the excellence of artists in their cinematic achievements, primarily in Hollywood or the American film industry. There are 25 categories for Golden Globes; 14 in motion pictures and 11 in television. At present, the Academy Awards has 24 categories. Usually, the Golden Globes ceremony is held in January of each year while the Academy Awards ceremony is held in February of each year.  \\ \\
            \# User question: Did any of my favorite actresses win any of them?  \\ \\
            \# Generated queries: \\ \\
            \hspace{10pt}1. Has Jennifer Aniston ever won a Golden Globe or an Academy Award?  \\
            \hspace{10pt}2. Has Lisa Kudrow ever won a Golden Globe or an Academy Award? \\
            \hspace{10pt}3. Did Jennifer Aniston win any awards for 'The Morning Show'? \\
            \hspace{10pt}4. List of Golden Globe winners in Best Actress in a Television Series – Musical or Comedy category for 1998 and 2003. \\ \\
            \hdashline \\
            \#  Example 2\\
            \# Background knowledge: \{$ptkb$\}\\ 
            \# Context: \{$ctx$\}\\
            \# User question: \{$user\ utterance $\} \\
            \# Generated queries: \\ \\
            \bottomrule
        \end{tabularx}

        \label{tbl:prompt-qd-llama}
 \end{table*}

\end{document}